# Local Electricity Market Design Utilizing Dynamic Network Usage Tariff

Bence Sütő, Dániel Divényi

## Abstract

The new technologies emerging in the energy sector pose new requirements for both the regulation and operation of the electricity grid. Revised tariff structures and the introduction of local markets are two approaches that could tackle the issues resulting from the increasing number of active end-users. However, a smooth transition from the traditional schemes is critical, thus creating the need for architecture that can be implemented in the current circumstances. This paper proposes a local market concept and a corresponding dynamic tariff system, which can be operated parallel to the current retail market. The participants of the market can trade energy peer-to-peer via a platform that allocates proper network charges to all transactions. The calculated tariffs consider the physical effect of the transactions on the grid in terms of nodal voltage deviations, branch current flows, and overall system losses. The proposed method is tested on the IEEE European LV test feeder through market simulations. The results imply that with the proper tuning of DNUT (Dynamic Network Usage Tariff) components, the end-users can realize surplus, while the security of network operation is also ensured.

## Nomenclature

| | | | |
|---|---|---|---|
| $p_d$ | price of demand order (EUR/MWh) | $I_{l,rated}$ | current rating of line "$l$" (A) |
| $p_s$ | price of supply order (EUR/MWh) | $U_n$ | phase RMS voltage on the node "$n$" (V) |
| $q_d$ | quantity of demand order (MWh) | $U_{n,nom}$ | nominal phase RMS voltage on the node "$n$" (V) |
| $q_s$ | quantity of supply order (MWh) | $I_{bt,l}$ | current RMS value before transaction on the line "$l$" (A) |
| $\Delta E_t$ | transacted energy (MWh) | $I_{at,l}$ | current RMS value after transaction on the line "$l$" (A) |
| $np$ | number of prosumers | $U_{bt,n}$ | phase RMS voltage before transaction on the node "$n$" (V) |
| $nn$ | number of nodes | $U_{at,n}$ | phase RMS voltage after transaction on the node "$n$" (V) |
| $nb$ | number of branches | $U_{neut,n}$ | neutral RMS voltage before on the node "$n$" (V) |
| $I_{branches}$ | current RMS values on each line (vector of size $nb$) | $lt_l$ | length of the line "$l$" (m) |
| $U_{nodes}$ | voltage phase RMS values on each node (vector of size $nn$) | $R_{line,l}$ | resistance of the line "$l$" (Ω) |

All Authors are at the Department of Electric Power Engineering, Budapest University of Technology and Economics, Budapest, Hungary (e-mail: suto.bence@vet.bme.hu).

| | | | |
|---|---|---|---|
| $\Delta E_{inj}$ | injected energy for every prosumer node (vector of size $3np$) | $R_{earth,n}$ | earthing resistance on the node "$n$" (Ω) |
| $I_l$ | current RMS value on the line "$l$" (A) | $c_{I,limit}$ | limiting current charge (EUR) |
| $IC_{limit,l}$ | cost of reaching the current limit on the line "$l$" (EUR) | $c_{I,linear}$ | linear current charge (EUR/Am) |
| $IC_{linear,l}$ | cost of current deviation on the line "$l$" (EUR) | $c_{U,limit}$ | limiting voltage charge (EUR) |
| $UC_{limit,n}$ | cost of reaching the voltage limit on the node "$n$" (EUR) | $c_{U,linear}$ | linear voltage charge (EUR/V) |
| $UC_{linear,n}$ | cost of phase voltage deviation on the node "$n$" (EUR) | $c_{loss}$ | linear loss charge (EUR/MWh) |
| $PC$ | cost of loss deviation (EUR) | EFR | excessive flow ratio |
| $DNUT_{s,d}$ | the DNUT value for a transaction between "$s$" supplier and "$d$" consumer (EUR/MWh) | PAR | prosumer activity ratio |
| PPR | prosumer participation ratio | SR | generator ratio compared to $np$ |

# 1 Introduction

The electric power system is subject to radical changes that are driven by the disruptive advancements in technology, such as household generators, storages, or electric vehicles. However, power grids and regulators are slow to respond to these novelties. One issue, in particular, is the state of distribution systems, since most of the new technologies can be tied to the end-users or so-called prosumers. Currently, these participants of the grid are considered as passive actors, but they become more and more active, thus generating a need to rethink traditional retail market structures and accompanying network tariffs, such as volumetric and capacity based measures [1].

We aim to find a solution, in which prosumers can trade energy between each other, thus gaining economic benefits. The constraints posed by the physical network (which may change in a short time) should also be taken into consideration by redefining network usage tariffs. Moreover, transactions that improve grid stability and energy quality should be incentivized. Nonetheless, the created structure should be prepared to be implementable in the regulations and operational circumstances present in the European Union.

In the context of these expectations, we identified three main approaches in the literature (often present at the same time) that address the aspects above:

- new tariff structures for retail markets,
- locational marginal pricing for distribution systems,
- creating local markets (or in a special case: peer-to-peer markets), which enable prosumers to trade locally generated energy with each other.

On the retail market, prosumers pay energy prices and network charges to separate companies. The main goal of tariff design is to incentivize consumers to shift their load according to the current state of the network. A considerable amount of papers have addressed this topic, from which we only mention two: [2] proposes a simple time-of-use (ToU) tariff, named "peak coincidence network charge", while [3] provides a game-theoretic approach to peak reduction by utilizing a flat-rate tariff. Although these tariffs mean an improvement from the operational perspective, the concept of cost-causality and dynamic behavior is missing.

Another approach to consider and quantify network use is if energy and grid prices are handled jointly. The distribution locational marginal pricing (DLMP) is a method, which distinguishes prosumers based on the state of the network. The DLMP can be derived from a standard optimal power flow (OPF) problem [4][5]. In the literature, there are several options to upgrade this method. One direction of development is to ensure that the OPF has a solution. [6] tackles this by using quadratic programming, [7] introduces unbalanced AC OPF, [8] linearizes the OPF problem, while [9] presents a relaxation method named convexified AC OPF. Other directions include utilization of the connection with the wholesale energy market [10], consideration of nodal voltages [11] and reactive power flows [8], or the application of stochastic models for renewable generators [12].

The DLMP is a promising approach for the reinvention of distribution system operation. However, since the method is based on optimization, active involvement of an operator (e.g. the DSO) is needed for the local trading to be successful. Another issue is that the concept of locational marginal pricing is yet to be accepted by EU regulators, thus making the implementation harder.

The concept and main attributes of local markets are summarized in [13] and [14]. The main idea is to create the opportunity for prosumers to trade with each other either through a central entity, like in [15], based on optimization, or peer-to-peer (p2p). P2p markets have seen a large interest in research over the years that span across various fields of science [16]. The results in this topic include the incorporation of blockchain technology [17], the utilization of game theory to describe trading [18][19], or to create models for prosumer behavior [20]. Others discuss providing incentives for self-consumption [21], and the introduction of network constraints in market operation [22], [23], which is a focus area in the present study, too.

P2p markets have the tools to incentivize end-users to operate in a market environment. Also, these structures can be easily demonstrated even with passive consumers. A barrier to implementation is the lack of thought on the parallel operation to the retail market.

Our contributions to local markets and dynamic tariffs are the following:

- We introduce a low-voltage (LV) local market platform, which can be operated in a parallel manner with the traditional retail market.
- We build a novel dynamic network usage tariff system (DNUT) that maps the state of the network to the market platform. This tariff is allocated on prosumers based on the effects of the transaction in terms of nodal voltages, branch currents, and system losses.
- The estimated state of the distribution network can be considered in the determination of the tariff components.

The operation of the local market platform and the attributes of the DNUT components are showcased through market simulations, using the IEEE European LV test feeder from [24].

This paper is organized as follows. Section II describes the local market concept and the incorporated calculation methods in detail. The attributes of the proposed market scheme and tariff structure are evaluated through market simulations in Section III. Finally, Section IV concludes this paper and provides insight into the continuation of this research.

# 2 Model

## 2.1 Local market concept

The proposed local market has similar attributes to the intraday wholesale electricity market. Trading is continuous to bridge possible liquidity issues. Quarter-hourly (QH) products are used, where a minimum limit for traded energy is 1 Wh.

One key feature of the local market structure is that it can work in parallel with the traditional retail market. This means that participation of prosumers is solely voluntary and is driven by profit-maximizing behavior. However, an accepted or hit order in the local market forms an obligation for both parties in the trade. The settlement is based on measurement in the given QH.

A trading platform is developed to connect all participants of the market, which might include:

- consumers (prosumers),
- small-scale generators,
- storage owners.

The platform is responsible for handling order submission and matching as well, as for settlement after physical delivery. Both supply and demand orders are composed of energy price and quantity. Dynamic network usage tariff charges are calculated for every pair of participants considering the actual state of the network. This fee is added to the price of the order and is shown on the market platform so that other prosumers cannot decompose the whole price into these two parts.

The DNUT depends on a base case, which is the forecasted state of the system (power generation and consumption data) for the actual QH. The corresponding DSO plays a key role in defining the DNUT components so that the resulting charges match the DSO's expectations in terms of how much it wishes to incentivize the grid supporting measures of participants (e.g. voltage regulation, alleviation of congestions).

A more detailed description of operation is given in the following subsections, starting with the calculation methods, developed for the local market.

## 2.2 Calculation methods

### 2.2.1 Sensitivity matrices

The DNUT considers the state of the network, which implies a large number of load-flow calculations to be carried out on the platform throughout the intraday operation. Therefore, sensitivity matrices are introduced to place the computational burden of load-flows before the trading period. This is done by examining voltage and current deviations from the aforementioned base case, as a function of energy exchange between each prosumer and the main grid (i.e. the slack bus). Through linearization, a voltage and current sensitivity factor (VSF, CSF) can be gained, which describes the networks' response to changes in energy injections and drains at prosumer nodes caused by transactions. These factors are ordered to obtain a VSF matrix of size $nn \times 3np$, and a CSF matrix of size $nb \times 3np$. The matrices have $3np$ columns to be able to consider the effect of single-phase prosumers too.

The matrices can be used to estimate the current state of the grid relative to the predefined base case (indexed with $bc$) by using the following equations:

$$U_{nodes} = U_{nodes,bc} + VSF \cdot \Delta E_{inj} \quad (1)$$

$$I_{branches} = I_{braches,bc} + CSF \cdot \Delta E_{inj} \quad (2)$$

Fig. 1. demonstrates the application of VSF and CSF parameters in a small LV network with 4 three-phase symmetrical prosumers. A base case is defined, where the prosumer on node 1 (N1) is a producer, and other participants on the network are consumers. The nodal voltages and branch currents for the calculated base case are denoted with blue. As mentioned earlier, the VSF and CSF values assume a "transaction" with the main grid. Therefore, the effect of a 250 Wh (1 kW for a QH) prosumer-prosumer transaction (red) is estimated as the sum of two parts: the producer supplies power to the main grid (yellow) and the consumer procures energy from the main grid (brown).

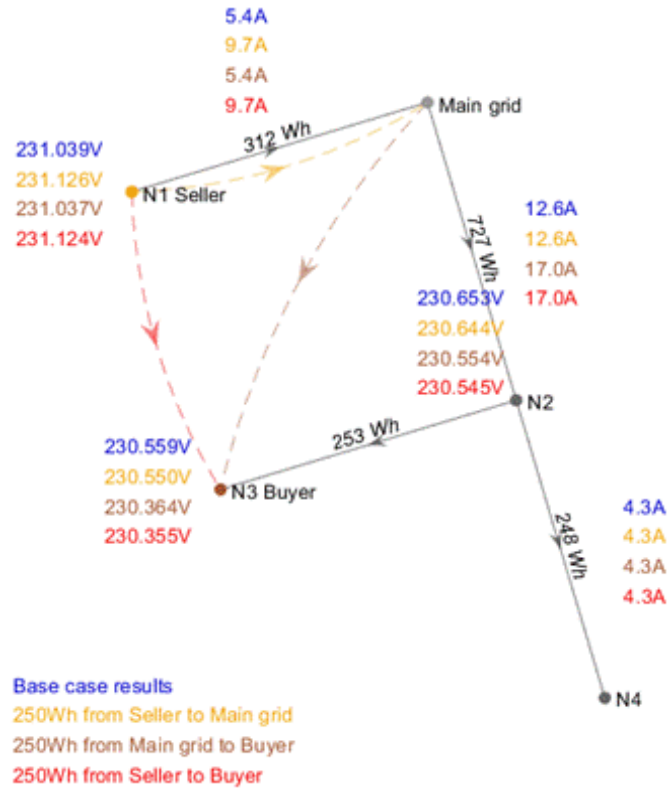

**Fig. 1.** Demonstration of the use of VSF and CSF matrices in a small network for a 250 Wh transaction between two prosumers

### 2.2.2 Dynamic Network Usage Tariff

A dynamic network usage tariff structure is used to introduce the physical constraints posed by the distribution system to the market platform. When a transaction between participants is considered, the DNUT is also added to (or subtracted from) the energy price, thus representing the effects of the transaction to those responsible. This serves as an incentive to hit orders that are advantageous from the grid perspective, or hinder other orders that would move the network towards a congested state.

For every pair of prosumers (also differentiated by transaction directions) a DNUT value can be derived, thus creating a DNUT matrix with the size of the number of prosumers. The diagonal of this matrix is zero (by definition), which means that transactions between prosumers on the same network node do not use the network.

The designed DNUT is composed of three main components: current charges, voltage charges, and loss charges. A detailed description of each component is given in the following subsections.

### 2.2.2.1 Current charges

For every line in the network, a current cost is calculated, which consists of a constant and a linear part. The constant part is responsible for keeping branch currents under the given rated values.

$$IC_{limit,l} = \begin{cases} 0, & I_I < I_{I,rated} \\ c_{I,limit}, & I_I \geq I_{I,rated} \end{cases} \quad (3)$$

$$IC_{limit} = \sum_l IC_{limit,l} \quad (4)$$

Since currents represent the traffic on the electricity grid, the linear part is calculated according to load allocation on the network.

$$IC_{linear,l} = c_{I,linear} \cdot \Delta I_l \cdot lt_l \quad (5)$$

$$IC_{linear} = \sum_l IC_{linear,l} \quad (6)$$

The substitution for $\Delta I_l$ can be done based on Table 1.

**Table 1** The possible values of $\Delta I_l$ in eq. (5)

| $\Delta I_l$ | | $I_{bt,l}$ + | $I_{bt,l}$ − |
|---|---|---|---|
| $I_{at,l}$ | + | $I_{at,l} - I_{bt,l}$ | $\lvert I_{at,l} \rvert - \lvert I_{bt,l} \rvert$ |
|  | − | $\lvert I_{at,l} \rvert - \lvert I_{bt,l} \rvert$ | $-(I_{at,l} - I_{bt,l})$ |

#### 2.2.2.2 Voltage charges

For every node in the network, a voltage related cost is calculated, which consists of a constant and a linear part. The constant part is responsible for keeping nodal voltages in a predefined ±10% interval around the nominal voltage (usually 231 V for LV networks), to ensure the quality of service.

$$UC_{limit,n} == \begin{cases} 0, & 0.9 U_{n,nom} \leq |U_n| \leq 1.1 U_{n,nom} \\ c_{U,limit}, & else \end{cases} \quad (7)$$

$$UC_{limit} = \sum_n UC_{limit,n} \quad (8)$$

The linear part of the voltage charge is used to incentivize voltage stabilizing trades (i.e. transactions that move the voltage towards the nominal value).

$$UC_{linear,n} = c_{U,linear} \cdot \Delta U_n \quad (9)$$

$$UC_{linear} = \sum_n UC_{linear,n} \quad (10)$$

The substitution for ΔU_n can be done based on Table 2.

**Table 2** The possible values of $\Delta U_n$ in eq. (9)

| $\Delta U_n$ | | $U_{bt,n} \geq U_{n,nom}$ | $U_{bt,n} < U_{n,nom}$ |
|---|---|---|---|
| $U_{at,n}$ | $\geq U_{n,nom}$ | $U_{at,n} - U_{bt,n}$ | $U_{at,n} + U_{bt,n} - 2U_{n,nom}$ |
|  | $< U_{n,nom}$ | $2U_{n,nom} - U_{at,n} - U_{bt,n}$ | $-(U_{at,n} - U_{bt,n})$ |

#### 2.2.2.3 Loss charges

In the case of network losses, only a linear charge is applied. The losses are estimated as the sum of branch losses and earthing losses in the following form:

$$P_{loss} = \sum_l I_l^2 \cdot R_{line,l} + \sum_n \frac{U_{neut,n}^2}{R_{earth,n}} \tag{11}$$

$$PC = c_{loss} \cdot (P_{loss,at} - P_{loss,bt}) \tag{12}$$

### 2.2.2.4 DNUT matrix

The DNUT value for two participants is constructed as:

$$DNUT_{s,d} = \frac{IC_{limit} + IC_{linear} + UC_{limit} + UC_{linear} + PC}{\Delta E_t} \tag{13}$$

A DNUT matrix with a size of (np×np) can be established by applying (20) to every pair of prosumers. Since all the components are derived from linear equations, the calculation of this matrix is fast.

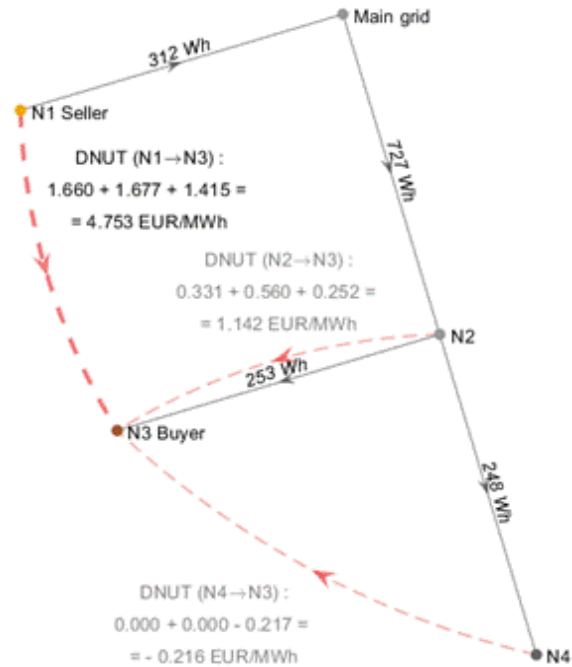

**Fig.2.** Demonstration of the DNUT values for the case when the consumer on node 3 buys 250 Wh energy

Fig. 2 shows an example of DNUT values in the same environment as in Fig. 1, but from the perspective of the prosumer on N3, who buys energy from the supplier on node 1. The tariff components from left to right are voltage charges, current charges, and loss charges. In this case, the flows resulting from the transaction match the direction of the base case flows, which means an additional burden for the grid (i.e. increased voltage drops, line currents and losses). Therefore, all DNUT components are positive.

If the buyer on N3 had the opportunity to buy energy from N4 (considering the same base case), the active power losses could be lowered on the branch between N2 and N4. This transaction would affect the voltages and currents in a symmetrical way, so that consequent costs cancel out each other. Thus, the result is a negative DNUT.

Note that an important part of the determination of DNUT is the tuning of the defined c constants. However, this topic is out of the scope of the current paper.

### 2.3 Order management

The order book management module is responsible for showing all available orders and corresponding DNUTs to participants. Furthermore, it handles the order list and logs the completed transactions, too.

Two orders can be paired if their type (either demand or supply) differs, and the following inequality is fulfilled.

$$p_d \geq p_s + DNUT_{s,d} \tag{14}$$

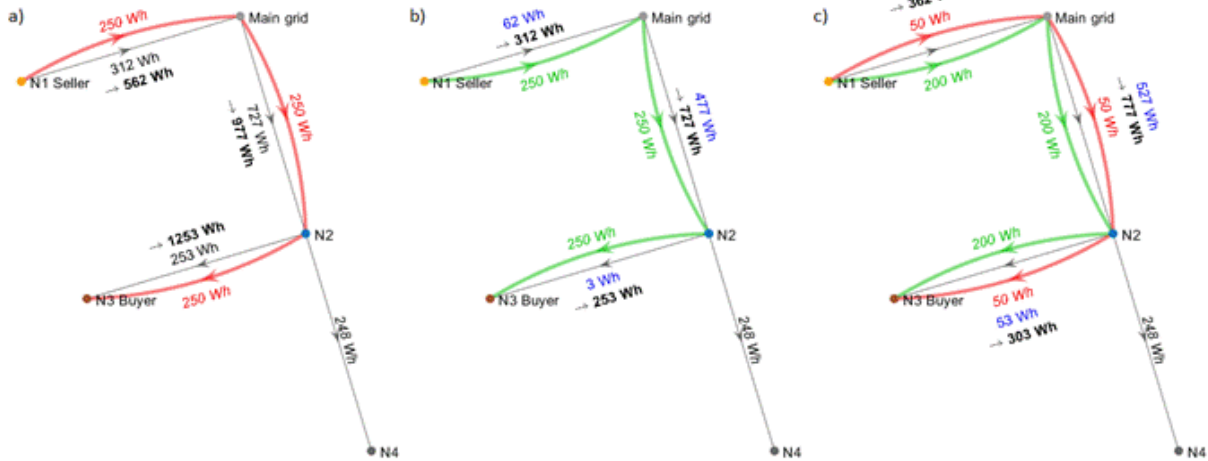

**Fig. 3.** Excessive (red), nominated (green), unnominated (blue), and resulting physical (bold) flows in the model

In the model, there are three ways to handle the power flow resulting from a transaction, which are described based on Fig. 3. The applied methods should be selected depending on the activity of prosumers.

In Fig. 3. a) the transaction is treated as an excessive flow (in red), that is added to the forecasted base case. This approach assumes active consumers who react to price signals on the platform (e.g. cheap energy generates more demand) resulting in new energy flow. Each trade creates a new system state, which will be the reference for further transactions. The DNUT matrix must be recalculated accordingly.

The method in Fig. 3. b) considers the transaction flow to be a part of the estimated base case, thus creating a nominated (green), and a remaining, unnominated (blue) flow. In this case, it is assumed that the market participants trade only their forecasted energy consumption/generation on the market platform to gain surplus. Each trade leaves the system state unchanged, and the initial DNUT matrix should not be updated. However, if a transaction exceeds the estimated base case, excessive flows are introduced similarly to Fig. 3. a).

The mixture of these two options is applied in Fig. 3. c), which is assumed to consider prosumer behavior more precisely. The ratio of nominated and excessive flows can be altered through the defined excessive flow ratio (EFR). In this example, the value of EFR is 0.2, which means that 80% of the transaction is nominated from the base case, while the remaining part is added to the network flows.

## 3 Results

The operation of the local market is presented through market simulations for one specific QH using the IEEE European LV test feeder (Fig. 4.). The network model is much more complex than the small test grid defined for Figures 1–3., as it contains the neutral line. 55 loads are present in the system, each of them connecting to one of the three phases, which introduces asymmetry in the simulations. An earthing resistance of 30 Ω is considered at prosumers.

## 3.1 General process and simulation parameters

The consumers are chosen according to a supply ratio (SR) to be switched to single-phase producers with a random generation volume taken from the [0,1] kWh interval. This enables trades on the local market, since without suppliers present in the system, only demand bids can be submitted.

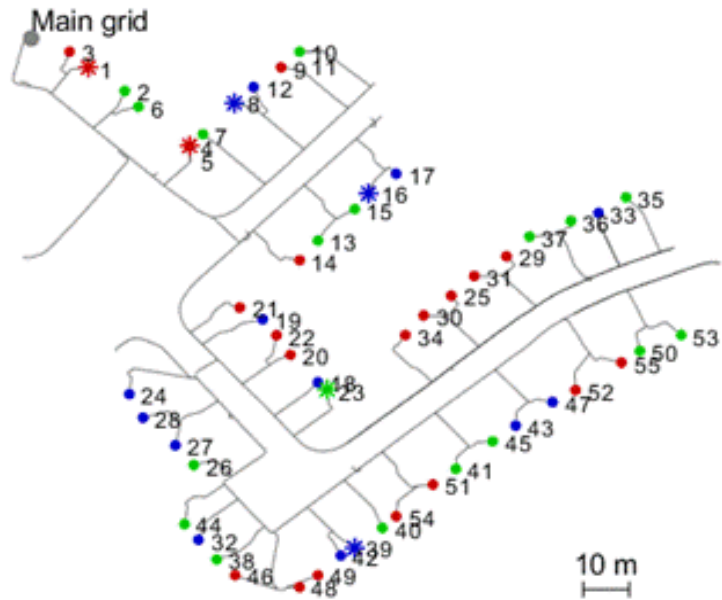

**Fig.4.** Network model with single-phase prosumers (red – phase a, green – phase b, blue – phase c, producers are marked by asterisks)

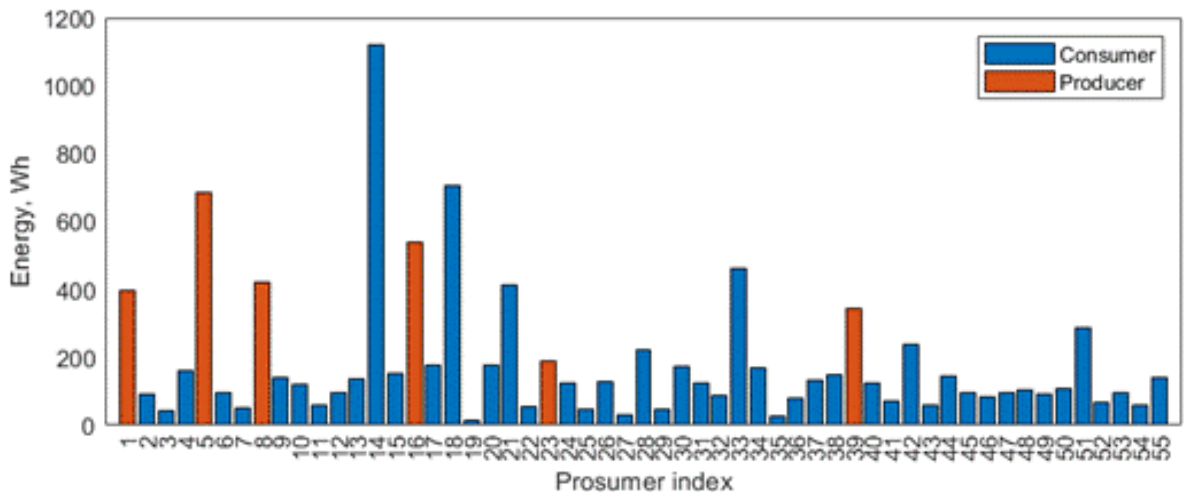

**Fig.5.** Produced (red) and consumed (blue) energy by prosumers in the given QH

In the market simulation, orders of the participants are created and matched automatically according to the following process:

1) A supply ratio is chosen from the [0,1] interval, which determines the number of producers on the network. The defined generators in the case of SR = 0.1 (6 in total) are marked in Fig. 4.
2) A participation ratio (PPR) is defined, which determines how many prosumers (out of 55) place orders. To ensure that supply-side orders are present, one producer is always selected to take part in the market. For example, a PPR of 10% means, that 5 consumers and 1 producer are active on the market.

3) An activity ratio (PAR) is applied, which determines how much energy these prosumers take to the market platform (through orders) relative to their base case power injections.
4) The prosumers are chosen in random order to place their order.
5) The order price is adjusted to the Hungarian retail price (approximately the same for every household consumer), which consists of a 39.03 EUR/MWh energy price, and 45.43 EUR/MWh distribution network usage tariff. It is assumed that every generator can sell energy for 39.03 EUR/MWh and every consumer can buy energy for 84.46 EUR/MWh in the retail market. Therefore, to make room for more profit, all orders' prices are generated randomly from the [39.03 84.46] EUR/MWh interval.
6) One step in the simulation implies the placement of exactly one order. This new order is either added to the order book or accepted if there is a suitable order in the book (that fulfills eq. (14)).

We mentioned earlier that the tuning of network charges is a complex problem, which requires further research. In this paper the costs are defined using the following method:

1) The initial DNUT matrix is calculated and decomposed into three parts: current, voltage, and loss tariff matrices.
2) The standard deviations of these component matrices are normalized.
3) Subsequently, the mean of these parameters can be adjusted to reach a given sum (15 EUR/MWh) of network usage tariff. This is assumed to be the task of the corresponding DSO.

The tunable parameters and their current values for the market simulations are summed in Table 3, while the base case energy injections are shown in Fig. 5.

The local market structure is tested along three dimensions, by changing SR, DNUT price constants, and PPR. During these studies, other parameters from Table 3 are held constant.

**Table 3** Tunable parameters and their values in the presented simulation cases

| Parameter | Cases I and III | Case II |
|---|---|---|
| $EFR$ | 0.2 | 0.2 |
| $PAR$ | 0.3 | 0.3 |
| $PPR$ | 0.3 | 0.3 |
| $SR$ | 0.1 | 0.1 |
| $c_{I,limit}$ (EUR) | 10 | 10 |
| $c_{I,linear}$ (EUR/A) | $3.588e^{-6}$ | $1.3952e^{-5}$ |
| $c_{U,limit}$ (EUR) | 10 | 10 |
| $c_{U,linear}$ (EUR/V) | $1.538e^{-4}$ | $2.3786e^{-4}$ |
| $c_{loss}$ (EUR/MWh) | 197.461 | 2047.99 |

### 3.2 Case I: The effect of changing supply side on DNUT

Through the course of this simulation, the value of SR is changed from 0.1 to 0.9 by 0.1 steps, meaning that with every iteration, additional 6 consumers are changed to be producers. Consequently, the base case, and thus the sensitivity matrices and the elements of DNUT matrix are reevaluated considering 0.4kWh trade between prosumers in each step.

Figure 6. shows the DNUT values from the perspective of Consumer 19 (connecting to phase c) to 3 suppliers (each of them connecting to different phases), respectively.

The two ends of Fig. 6. share similar attributes. When the number of generators is low, the direction of network flows in all three phases points away from the main grid towards consumers. The opposite is true in the case of a high supply ratio. In both scenarios, trades that alleviate the most congestions are incentivized the most. For example, at SR = 0.9 the transaction between Prosumers 19 and 1 causes a burdening flow between 1 and the main grid, and a relieving flow between the main grid and 19.

The simulation results for middle points (around SR = 0.5) depend on the process of simulation itself, i.e. generated base case energy injections, and the order, in which the consumers are switched to producers.

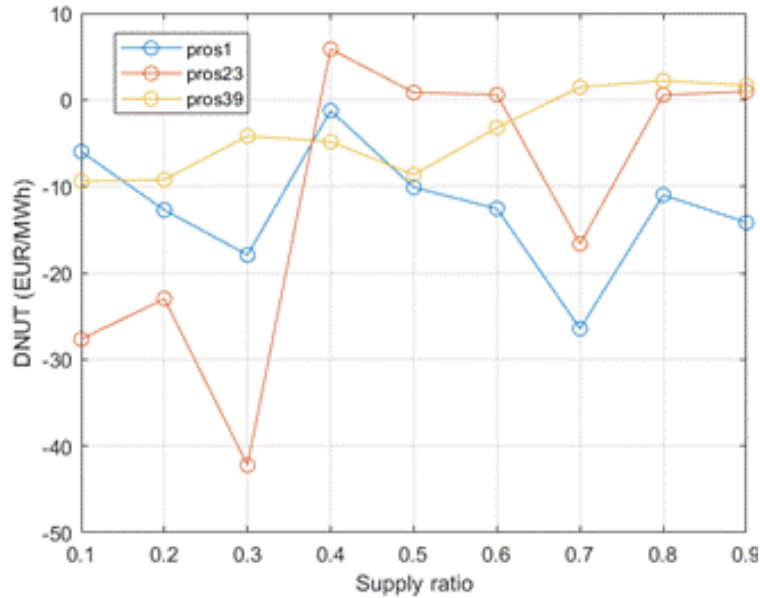

**Fig.6.** DNUT values between Prosumer 19 (consumer) and Prosumers 1, 23, 39 (generators on phases a-b-c respectively)

### 3.3 Case II: The effect of DNUT components on the market

In this case, market simulations are carried out using only the DNUT component(s) related to one physical parameter at the time (e.g. $c_{I,limit}$ and $c_{I,linear}$ are applied, while other prices are set to zero). The parameters are initialized so that the mean of the resulting DNUT matrix equals the mean of the original DNUT matrix (calculated according to the parameters in the second column of Table 3). This can be reached by using the constants of the third column of Table 3. In the simulations, these parameters (thus the mean of the DNUT matrix) are changed from 0.1 to 2 times the calculated value in 0.1 steps.

The results in the form of the surplus of participants relative to the number of active prosumers are shown in Fig. 7. The surplus of a participant is based on how much the price realized on the local market (energy price and DNUT) differs from the respective retail price. The social welfare gain in the given QH can be calculated as the sum of participants' surpluses. The division by the number of active prosumers is needed to ensure comparability between iterations. It can be noted that both in Fig. 7. and later in Fig. 8. the value of relative surplus is in the 0.0001 – 0.001 EUR interval. This is due to the low volumes traded on the local market (and generally at the distribution level).

Fig. 7. shows linearly increasing surplus, which is due to predominantly negative DNUT values. Other than that, the biggest effect on surplus stems from loss charges, followed by current charges and voltage charges. These relations might turn around, but this type of simulation can help in the tuning of DNUT components in the operation planning phase.

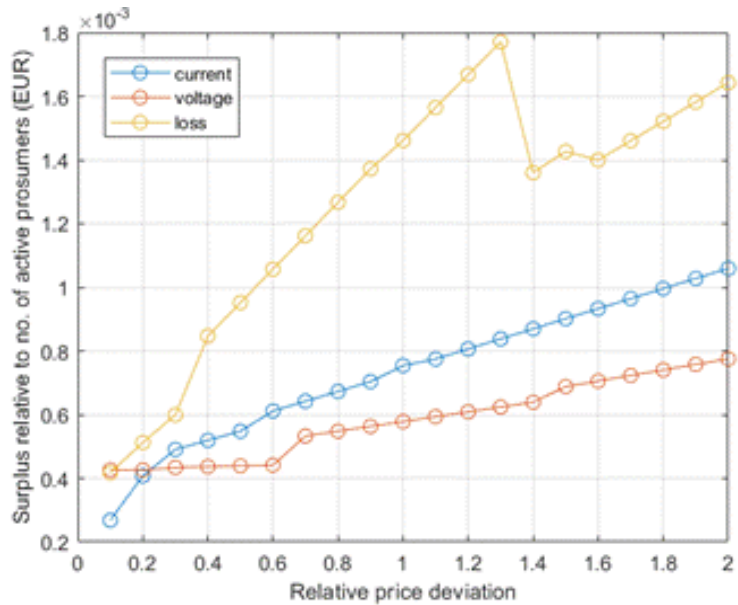

**Fig.7.** The change of surplus in the system as a result of increasing DNUT components applied separately

## 3.4 Case III: Effects of market activity

At this stage of the study, the effect of growing PPR (from 0 to 1 by 0.1 steps) is evaluated through a Monte Carlo simulation. A single market simulation is carried out 50 times, using a different set of orders. The participating prosumers submit exactly one order in each iteration. The change of the surplus over the course of the simulation is depicted in Fig. 8.

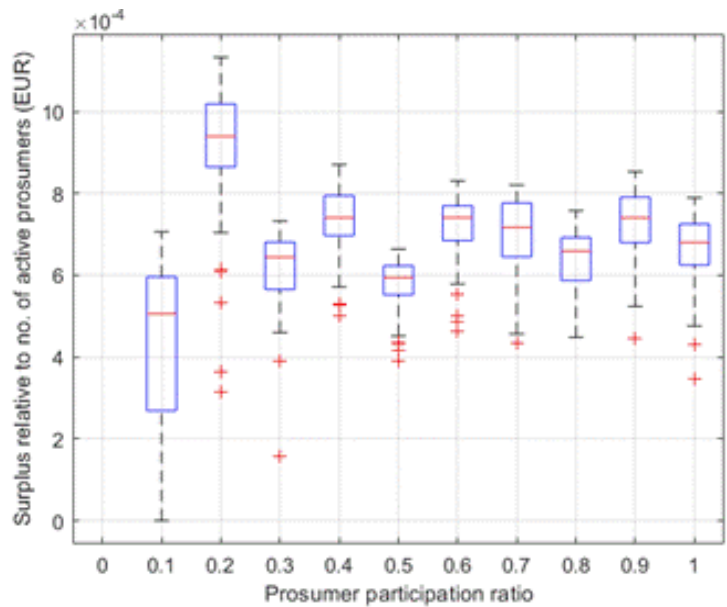

**Fig.8.** The change of surplus in the system as a result of increasing PPR in Monte Carlo simulation

In general, the surplus increases with growing participation ration, which results in an approximately constant relative surplus curve. Small oscillations are superimposed over this constant, which is due to the ever-changing ratio of active producers to active consumers. At low PPR values (0.1, 0.2), this ratio is 1:5, thus the increase in relative surplus comes from the fact that generators are forecasted to inject more energy into the network compared to consumers. Therefore, more orders of the demand side can be paired with the same supply order.

Fig. 9. shows the mean of DNUT costs for all transactions in the actual QH, decomposed into voltage related, current related, and loss related parts.

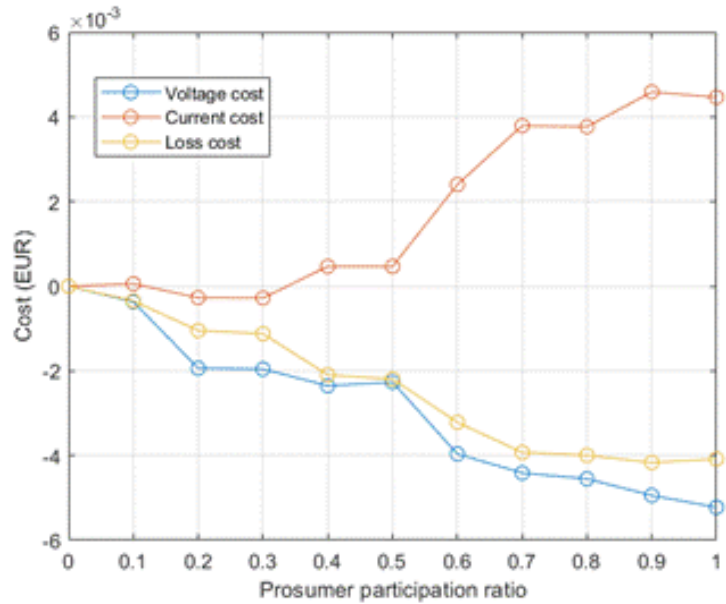

**Fig.9.** The mean of sum DNUT costs decomposed by physical measures as a result of increasing PPR in Monte Carlo simulation

The network is dominated by consumption, to which energy mainly comes from the main grid. Because of that, local trading is incentivized throughout the simulation, resulting in negative DNUT costs (i.e. participants are paid to use the local market platform). However, not all parts of the DNUT cost remain in the negative territory. The asymmetric topology of the network causes energy trades to use a big part of the grid (formerly discussed in section B); and therefore, raises current costs for active prosumers.

# 4 Conclusion

In this paper we have proposed a local market concept, in which prosumers can trade peer-to-peer, aided by a market platform. To assess the physical effects of trades on participants, the nodal voltages and the branch currents as a function of injected energy have been linearized, thus introducing sensitivity factors. Based on this computational method, a dynamic tariff structure has been developed and incorporated in the prices of orders. Simulations show that the DNUTs influence the market transactions in such a way that is beneficial from the network perspective: voltage and current limits have not been reached, and local generation (in a consumption-heavy area) has been incentivized.

A demonstration will be carried out, through which the market structure will be tested in three different LV sites. These tests are expected to uncover the unknown attributes and potential barriers to the system.

In our future work, the methodologies supporting the proposed local market scheme need to be developed. These include the precise method used for base case prediction and the algorithm for tuning tariff components to provide DNUT, which reflects the state of the network and the needs of the DSO.

# Acknowledgements

This work was partly supported by the European Union's Horizon 2020 research and innovation programme under grant agreement No. 824330.

The research reported in this paper and carried out at the BME has been supported in part by the NRDI Fund based on the charter of bolster issued by the NRDI Office under the auspices of the Ministry for Innovation and Technology.